\documentclass{aip-cp}
\usepackage[numbers]{natbib}
\usepackage{rotating}
\usepackage{graphicx}
\newcommand{\GeV}{\mathrm{GeV}}
\newcommand{\pipipi}{\pi^-\pi^+\pi^-}
\newcommand{\pipi}{\pi^+\pi^-}

\newcommand{\widthTwo}{.45\columnwidth}
\begin{document}

\title{The \boldmath$2\pi$ Subsystem in Diffractively Produced \boldmath$\pi^-\pi^+\pi^-$ at \textsc{Compass}}

\author[aff1]{Fabian Krinner\corref{cor1}}
\author[]{for the \textsc{Compass} collaboration}

\affil[aff1]{Technische Universit\"at M\"unchen, Physik-Department, E18}
\corresp[cor1]{Corresponding author: fabian-krinner@mytum.de}

\maketitle

\begin{abstract}
The \textsc{Compass} experiment at CERN has collected a large dataset of $50$ million $\pipipi$
events produced diffractively from a proton target using a $190\GeV/c$ pion beam. The partial-wave analysis
(PWA) of these high-precision data reveals previously unseen details but is limited in parts
by systematic effects.
The PWA is based on the isobar model, in which multi-particle decays are described
as a chain of subsequent two-body decays. Here, fixed mass distributions for the appearing
intermediate resonances, the so-called isobars, are assumed. These shapes, which e.g.
may be parametrized by Breit-Wigner amplitudes, represent prior knowledge that has to
be put into the analysis model and may therefore introduce a model dependence, thus
increasing systematic uncertainties. We present a novel method, which allows to extract
isobar amplitudes directly from the data in a more model-independent way. As a first
application, diffractively produced $\pipipi$ events are analyzed. Here, the focus lies in
particular on the scalar $\pipi$ subsystem, where in a previous analysis a signal for a new
axial-vector state $a_1(1420)$ was found in the $f_0(980)\pi$ decay mode.
\end{abstract}

\section{THE COMPASS EXPERIMENT}
\label{sec::compass}
The two-stage \textsc{Compass} spectrometer is located at CERN's Northern Area. It is supplied with various types of beam particles (e.g. secondary hadron or tertiary muon beams) by the Super Proton Synchrotron.
The different beams and target setups, in combination with a good acceptance over a wide kinematic range allows for a wide physics program studying the structure as well as the sprectrum of hadrons, 
they will be presented here.\\
For the analysis shown here, data from the year $2008$ are used. Which were recorded using a $190\GeV/c$ negative hadron beam, consisting of $97\%$ of $\pi^-$, which was directed onto a $40\,\mathrm{cm}$ long liquid hydrogen target. 
The rest of the beam consisted mostly of $K^-$ and antiprotons.

\section{PARTIAL-WAVE ANALYSIS METHOD}
\label{sec::pwa}
\subsection{The Isobar Model}
\label{sec::isobar}
To describe the process $\pi^-p\to X^-p\to\pi^-\pi^+\pi^-p$ one basic assumption is made. This assumption is called the isobar model, which states that the appearing intermediate $3\pi$ state $X^-$ does not decay directly into $\pipipi$, 
but undergoes 
subsequent two-particle decays until it ends up in the final state: $X^-\to\xi^0\pi\to\pipipi$.\\
The appearing intermediate two-pion state $\xi^0$, which is called the isobar, can be any known $\pipi$ resonance with the appropriate quantum numbers. In the considered channel there are e.g. $f_0(980)$, $\rho(770)$, or $f_2(1270)$.

\subsection{Conventional PWA}
\label{sec::conventional}
In a conventional PWA \cite{Adolph:2015tqa} the complex decay amplitude, describing the measured intensity distribution $\mathcal{I}$ is 
expanded in the following way:
\begin{equation}
\label{eq::intens}
\mathcal{I}(m_{3\pi}, t^\prime, \vec\tau) = \left| \sum_\mathrm{waves} T_\mathrm{wave}(m_{3\pi}, t^\prime) \mathcal{A}_\mathrm{wave}(\vec\tau) \right|^2,
\end{equation}
where the $T_\mathrm{wave}$ are the production amplitudes depending on the invariant mass $m_{3\pi}$ of the $\pipipi$ system and on the squared four-momentum transfer $t^\prime$. These production amplitudes are fitted to the data
in bins of the kinematic variables $m_{3\pi}$ and $t^\prime$, using a maximum likelihood fit.\\
The decay amplitudes $\mathcal{A}_\mathrm{wave}$, depending on the remaining $5$ kinematic variables represented by $\vec\tau$, are calculable and have to be put into the fit beforehand. They can be split further into a mass-dependent 
part $\Delta(m_{\pi^+\pi^-})$ which depends on the mass of the $\pipi$ subsystem, and an angle-dependent part $\Psi(\vec\Theta)$:
\begin{equation}
\mathcal{A}_\mathrm{wave}(\vec\tau) = \Delta(m_{\pipi})\Psi(\vec\Theta) + (\mathrm{Bose\ symmetrization\ term}).
\end{equation} 
The amplitude $\Psi(\vec\Theta)$ is determined by the angular momenta, while the mass-dependent amplitude $\Delta(m_{\pi^+\pi^-})$ has to be known without any free parameters from previous experiments to match the complex amplitude
 of the corresponding isobar. In the simplest case Breit-Wigner amplitudes are used.
The Bose symmetrization term is necessary due to the two indistinguishable $\pi^-$ in the final state.\\
The isobar mass-shape has to be known for this kind of analysis, but no unique parametrizations are given by theory. Therefore some bias may be introduced by selecting a particular parametrization. 
\subsection{A Novel Approach}
\label{sec::novel}
To remove this bias a novel approach was introduced, which was inspired by Ref. \cite{Aitala:2005yh}. It does not rely on previous knowledge on the isobar amplitude, but allows to extract it from the data. To this end, the parametrization of the isobar amplitude is replaced by a set of piece-wise 
constant complex functions, that introduce a binned $m_{\pi^+\pi^-}$ dependence. They are defined in the following way:
\begin{equation}
\Delta_\mathrm{bin}(m_{\pi^+\pi^-}) = \left\{\begin{array}{l} 1 \mathrm{\ if\ }m_{\pi^+\pi^-}\mathrm{\ lies\ in\ the\ corresponding\ mass\ bin}\\
							      0 \mathrm{\ otherwise}\end{array}\right.
\end{equation}
With this definition, the continuous isobar amplitudes are replaced according to:
\begin{equation}
\Delta(m_{\pi^+\pi^-}) \to \sum_\mathrm{bins} \Delta_\mathrm{bin}(m_{\pi^+\pi^-})
\end{equation}
Inserting this into equation (\ref{eq::intens}), the intensity reads:
\begin{equation}
 \mathcal{I}(m_{3\pi}, t^\prime, m_{\pi^+\pi^-}) = \left| \sum_\mathrm{waves} \sum_\mathrm{bins} T_\mathrm{wave}^\mathrm{bin}(m_{3\pi}, t^\prime, m_{\pi^+\pi^-})\left(\Delta_\mathrm{bin}(m_{\pi^+\pi^-})\Psi_\mathrm{wave}(\vec\Theta)\right)+(\mathrm{Bose\ symmetrization\ term})\right|^2
\end{equation}
The new piece-wise constant isobar amplitudes behave just like additional partial waves. The corresponding production amplitudes $T_\mathrm{wave}^\mathrm{bin}$ now also contain information about the binned isobar amplitude.

\section{RESULTS}
\label{sec::results}
\subsection{Data and Wave Set}
\begin{figure}[h]
\label{fig::comparison}
\includegraphics[width=.33\columnwidth]{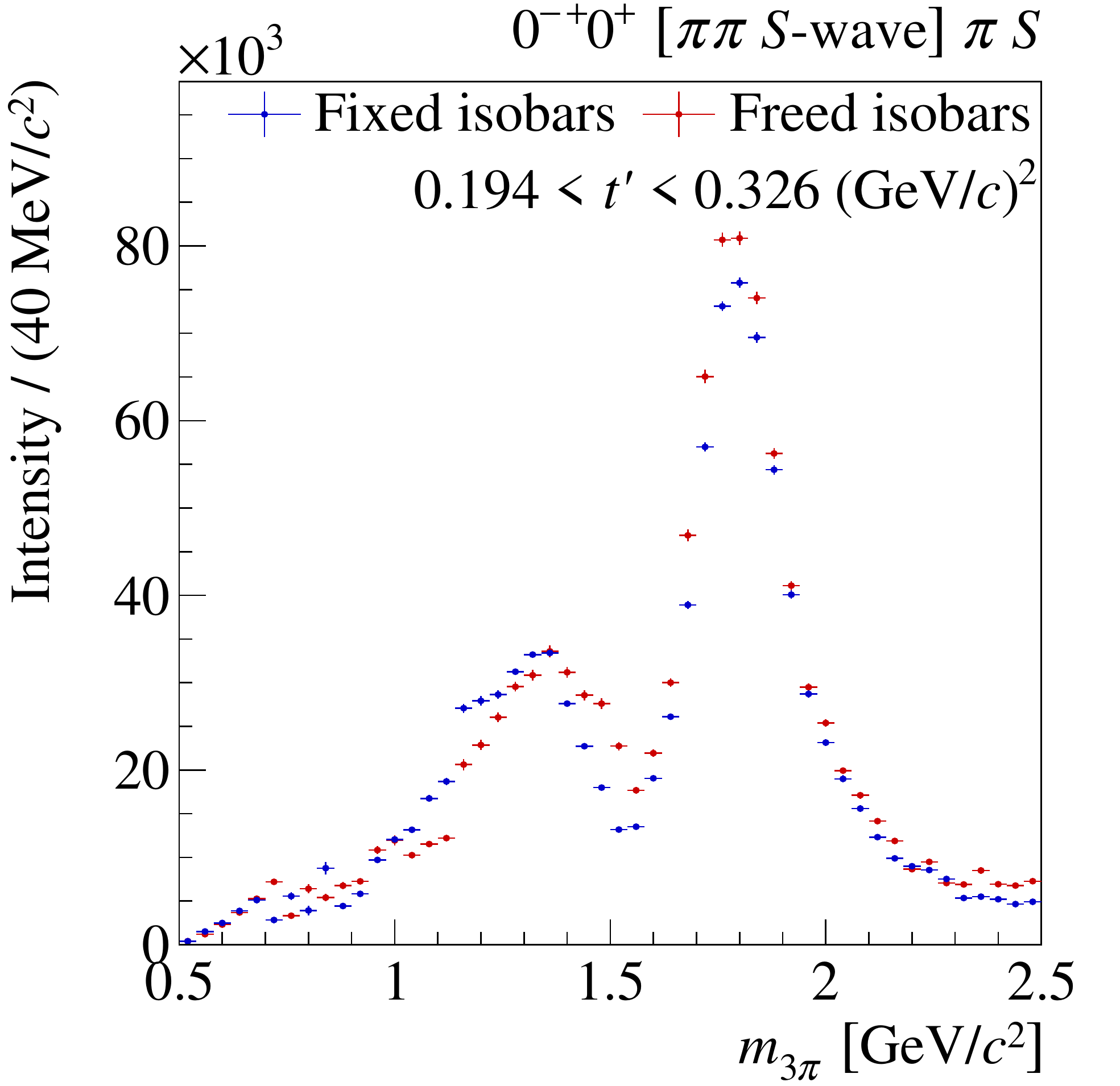}
\includegraphics[width=.33\columnwidth]{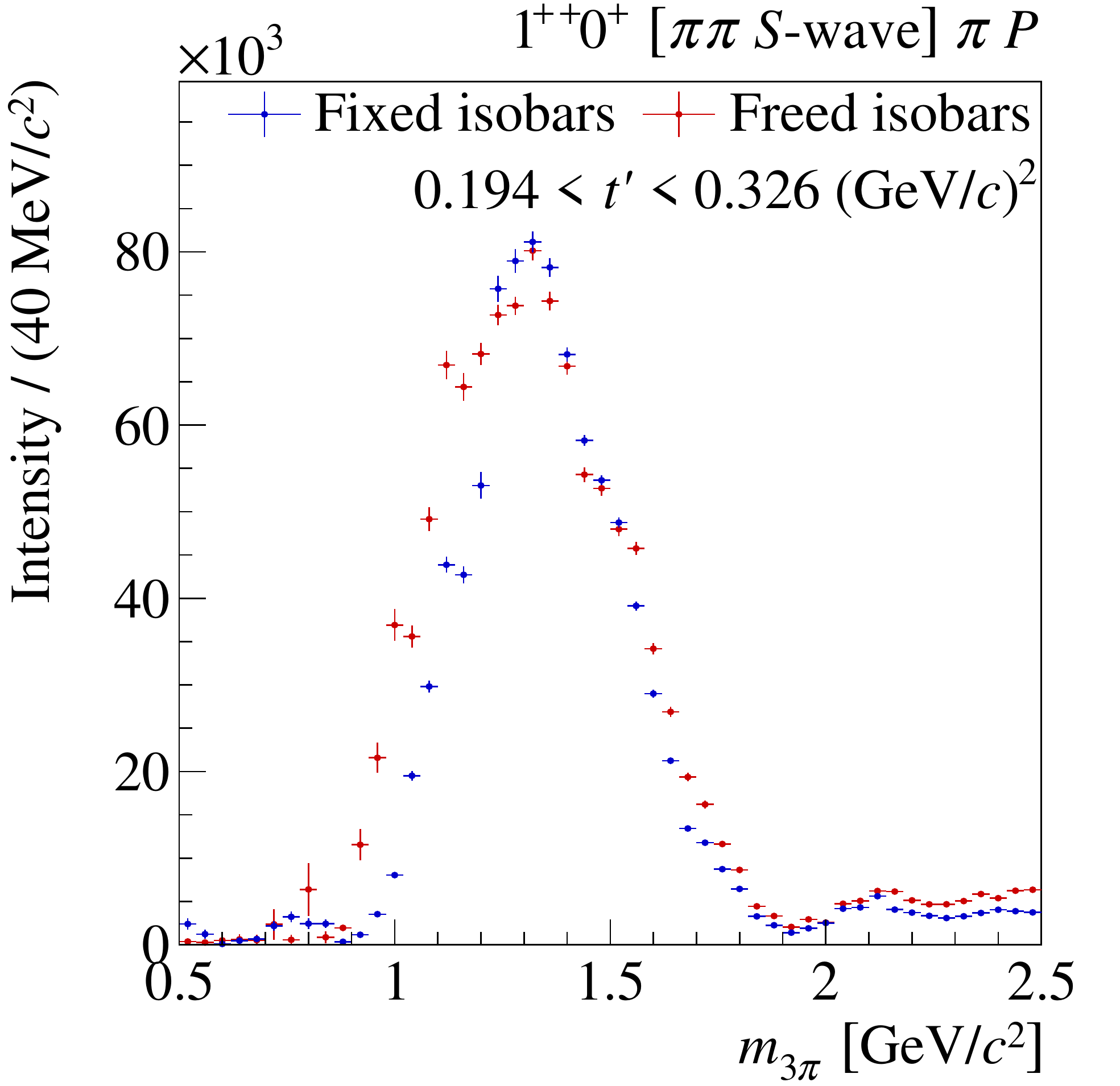}
\includegraphics[width=.33\columnwidth]{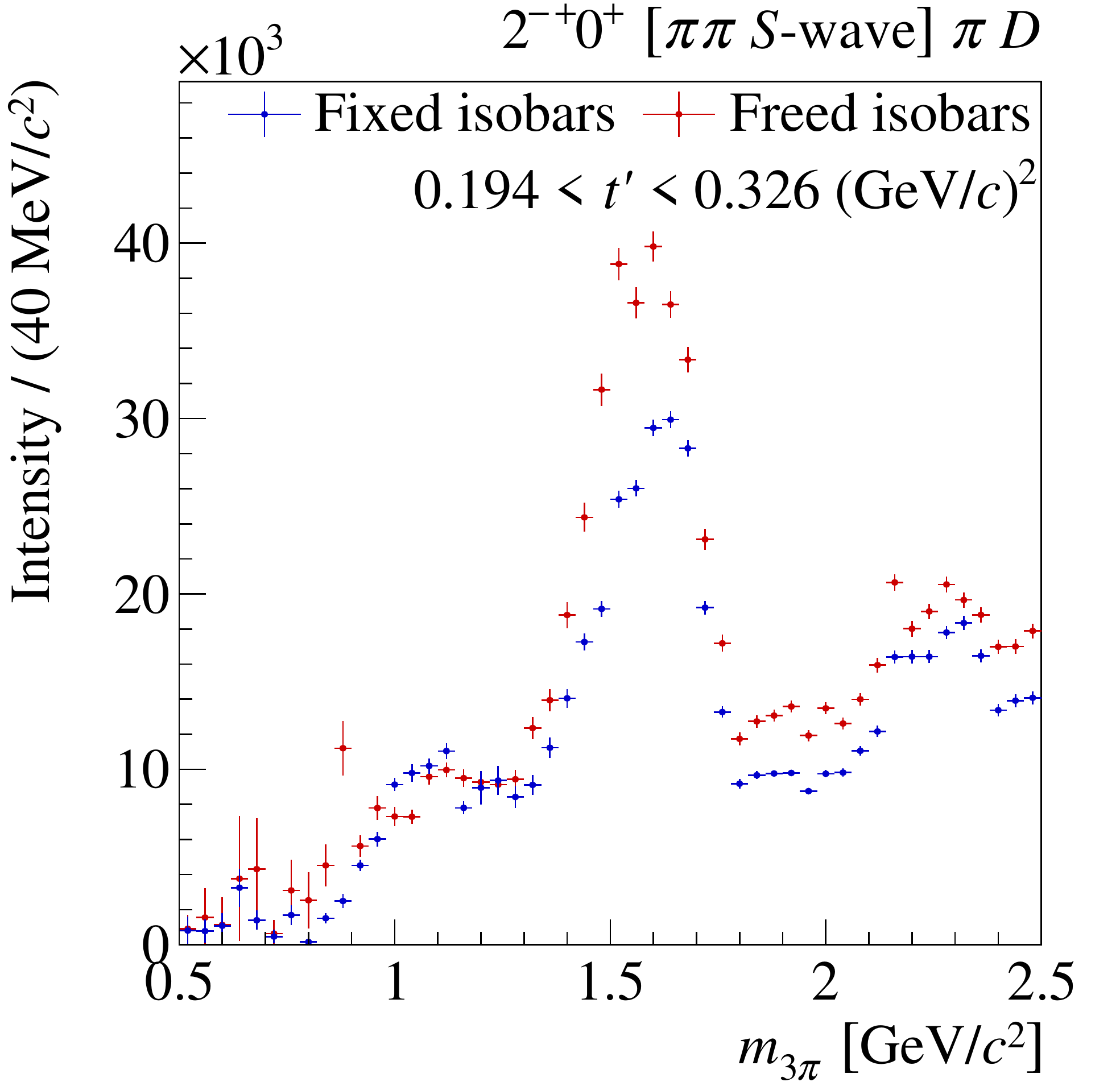}\\
\caption{Comparison of the intensity distributions for the $0^{-+}0^+[\pi\pi]_{0^{++}}\pi\,S$ wave (left), the $1^{++}0^+[\pi\pi]_{0^{++}}\pi\,P$ wave (middle) and the $2^{-+}0^+[\pi\pi]_{0^{++}}\pi\,D$ wave (right) obtained with conventional PWA (blue) and freed-isobar PWA (red) \cite{Adolph:2015pws}.}
\end{figure}

The freed-isobar approach described above was applied to the large data set collected with the \textsc{Compass} spectrometer for the process $\pi^-p\to\pi^-\pi^+\pi^-p$, which consists  approximately $50$ million events. The model
that was employed is based on a detailed PWA of this data, using a set of $88$ waves \cite{Adolph:2015tqa}.\\
In the analysis presented in the following, seven of these waves with $J^{PC} = 0^{++}$ isobars were replaced by sets of piece-wise constant functions:
\begin{eqnarray}
	\left.\begin{array}{l}
		0^{-+}0^+f_0(500)\pi\,S\\
		0^{-+}0^+f_0(980)\pi\,S\\
		0^{-+}0^+f_0(1500)\pi\,S
	\end{array}\right\} = 0^{-+}0^+[\pi\pi]_{0^{++}}\pi\,S \\ 
	\left.\begin{array}{l}
		1^{++}0^+f_0(500)\pi\,P\\
		1^{++}0^+f_0(980)\pi\,P\\
	\end{array}\right\} = 1^{++}0^+[\pi\pi]_{0^{++}}\pi\,P \\ 
	\left.\begin{array}{l}
		2^{-+}0^+f_0(500)\pi\,D\\
		2^{-+}0^+f_0(980)\pi\,D\\
	\end{array}\right\} = 2^{-+}0^+[\pi\pi]_{0^{++}}\pi\,D
\end{eqnarray}
Since these seven waves employ only three different combinations of angular momenta, only three sets of piece-wise constant functions are necessary to replace them.\\
Comparisons between conventional and novel PWA methods are shown in Fig. \ref{fig::comparison}, where the corresponding amplitudes are summed. The sum runs over all fixed isobars or $m_{\pipi}$ bins respectively. 
All three waves show a good overall agreement.
\subsection{\boldmath$0^{-+}0^+[\pi\pi]_{0^{++}}\pi\,S$}
\begin{figure}[h]
\label{fig::0mp}
\includegraphics[width=\widthTwo]{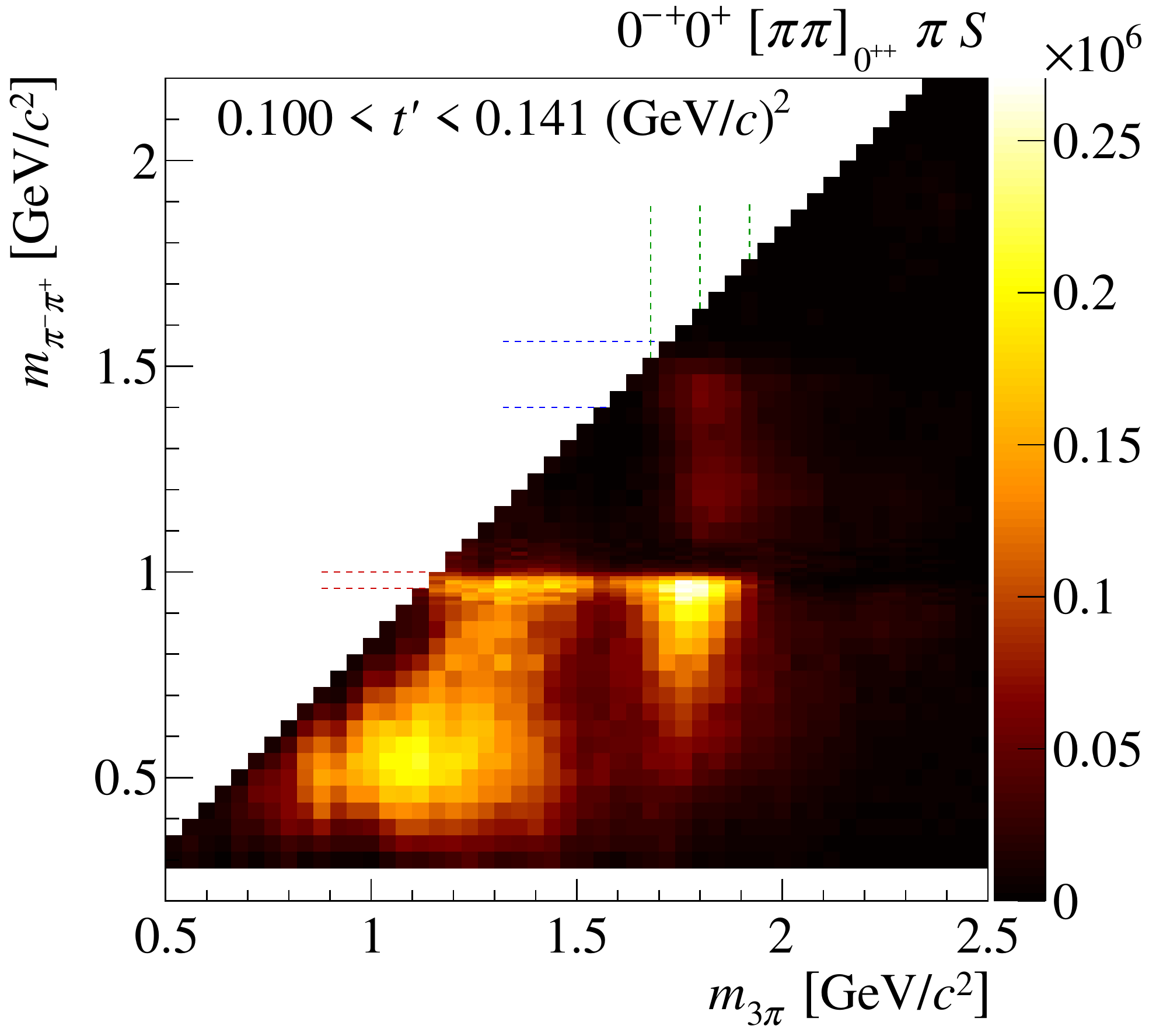}
\includegraphics[width=\widthTwo]{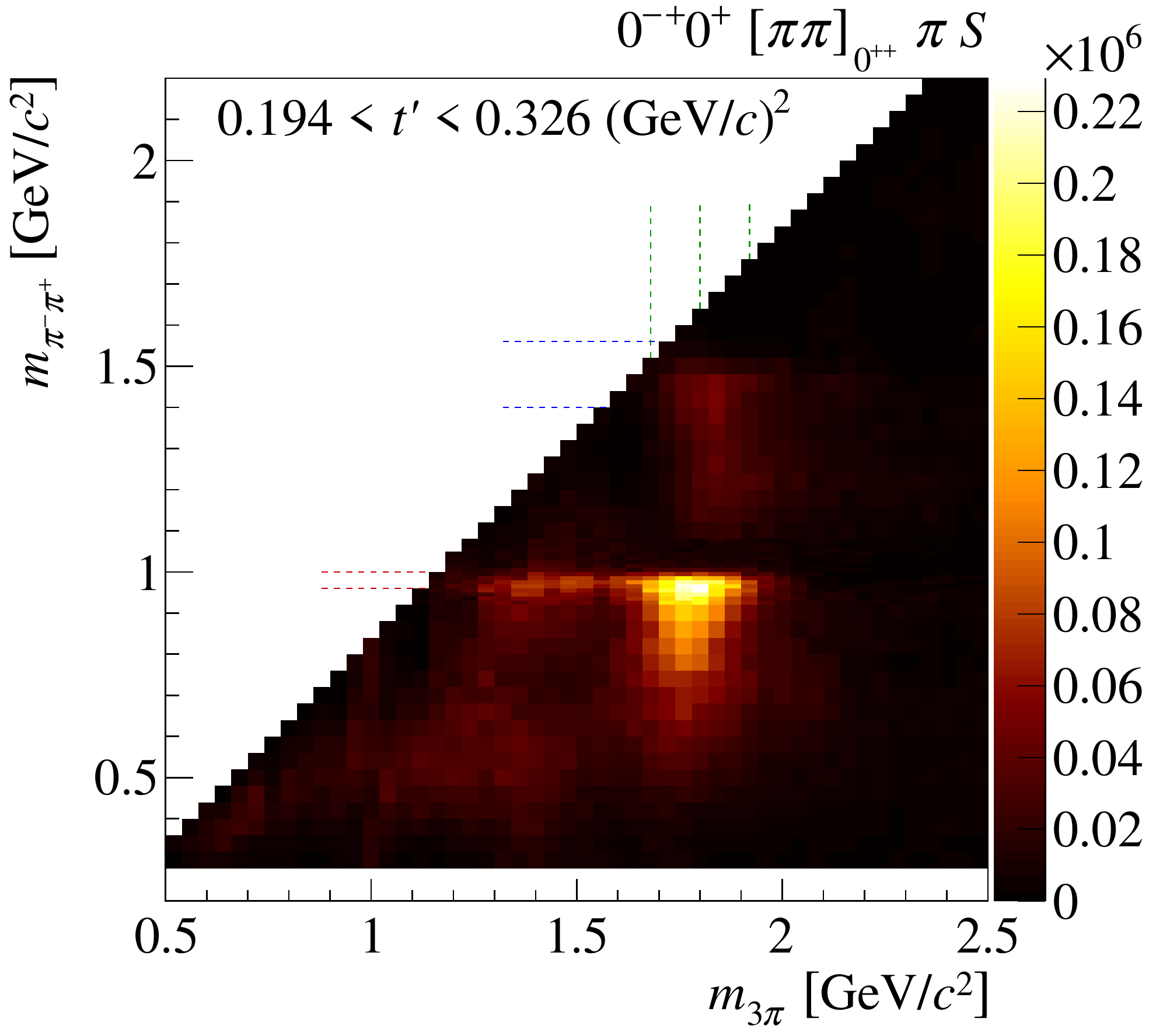}
\caption{Two-dimensional intensity distribution of the $0^{-+}0^+[\pi\pi]_{0^{++}}\pi\,S$ wave depending on $m_{3\pi}$ and $m_{\pipi}$ for two different bins of the four-momentum transfer $t^\prime$ \cite{Adolph:2015pws}.}\\
\end{figure}

\begin{figure}[h]
\label{fig::0mpInt}
\includegraphics[width=.33\columnwidth]{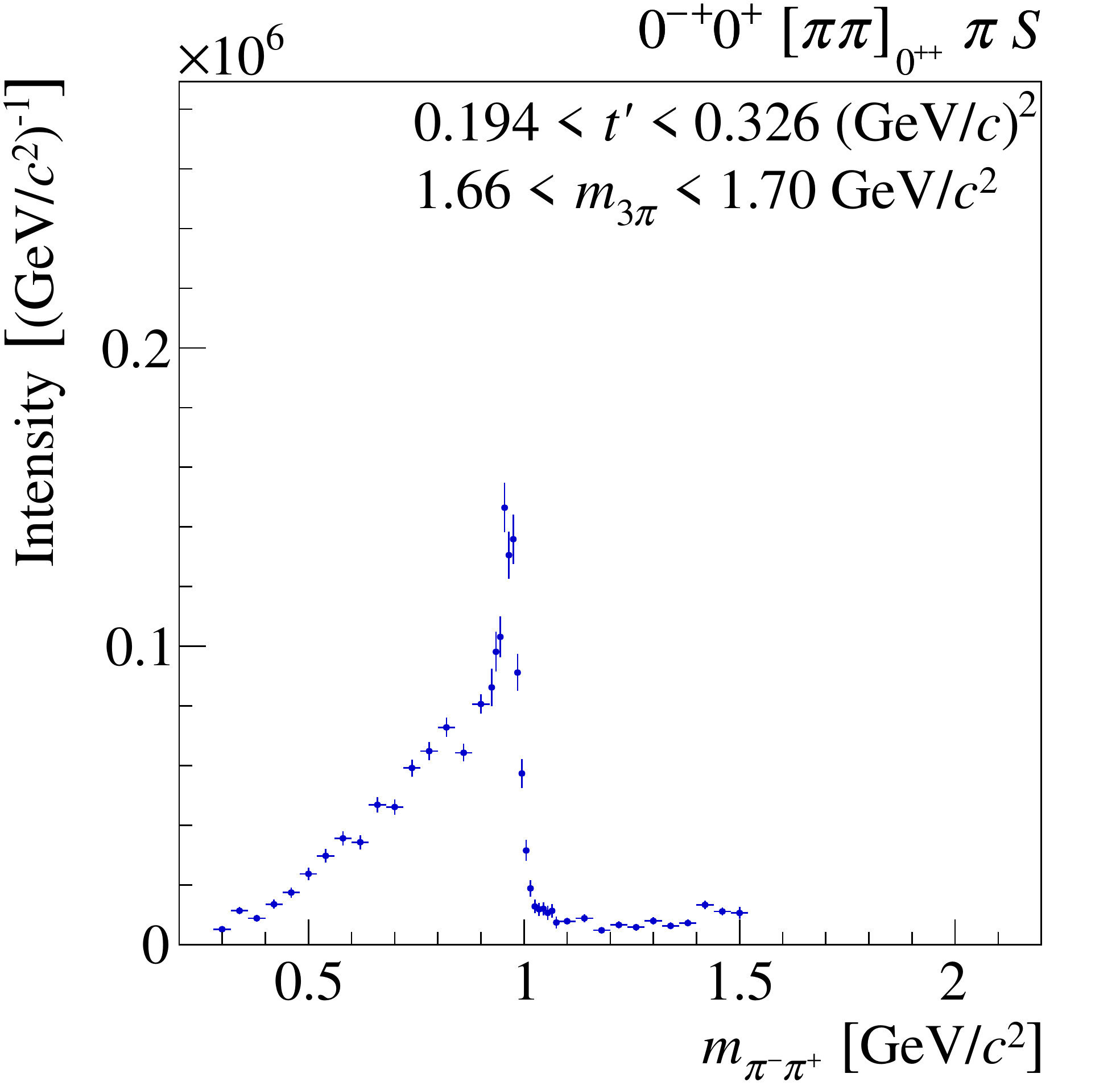}
\includegraphics[width=.33\columnwidth]{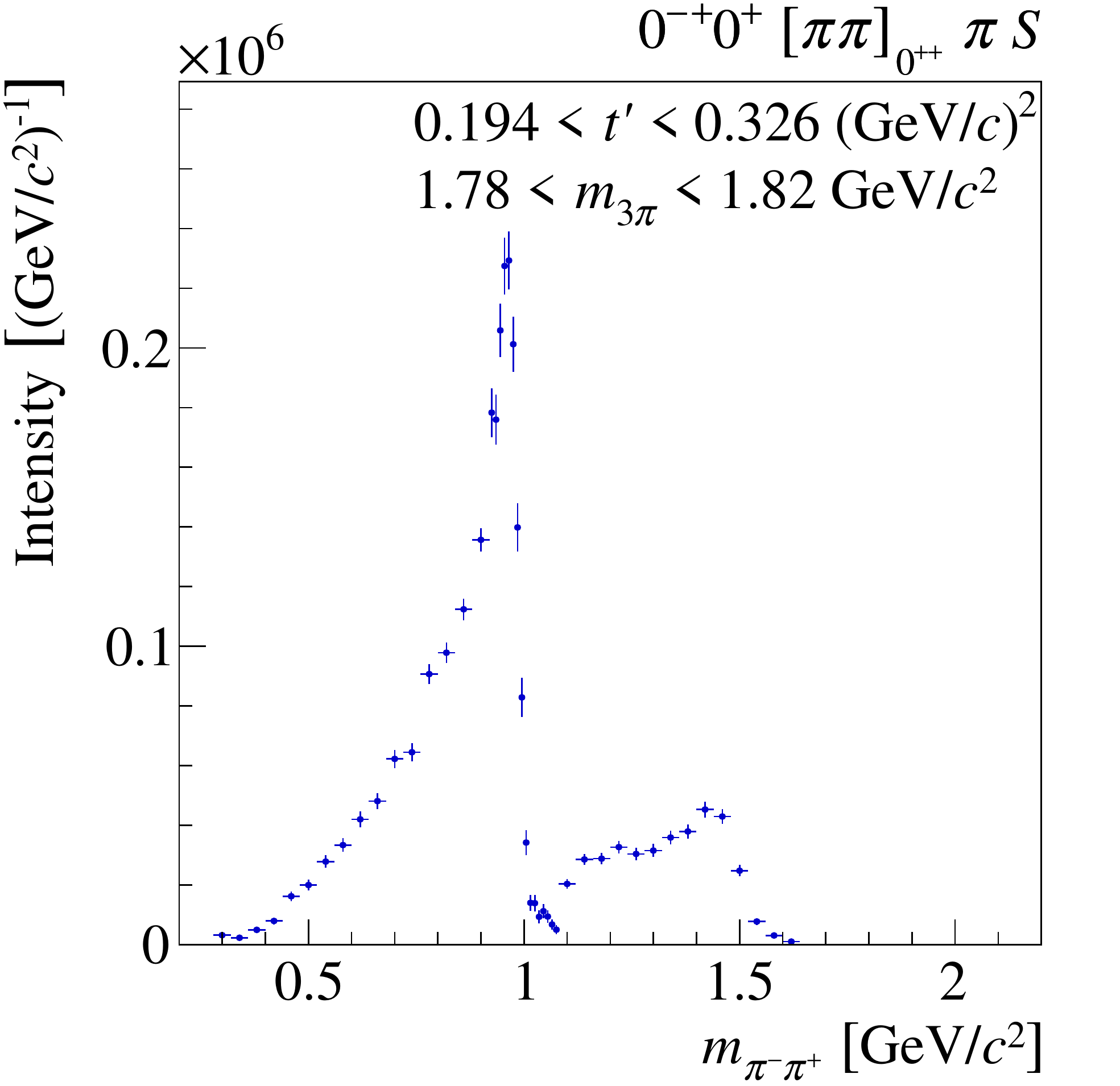}
\includegraphics[width=.33\columnwidth]{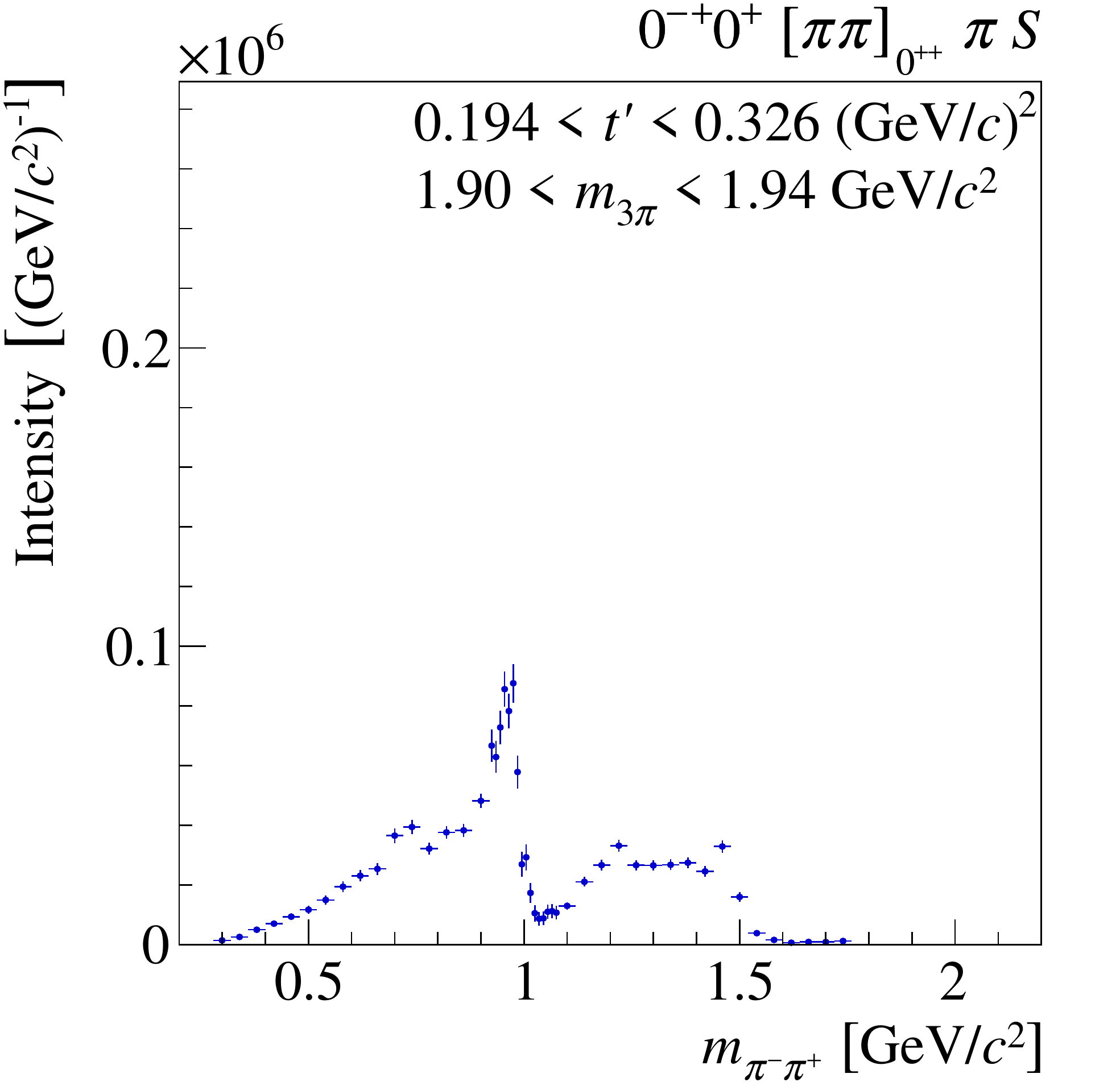}\\
\caption{Intensity distributions of the extracted $\pipi$ amplitude, below (left), on (middle) and above (right) the $\pi(1800)$ resonance \cite{Adolph:2015pws}.}
\end{figure}

\begin{figure}[h]
\label{fig::0mpArg}
\includegraphics[width=.33\columnwidth]{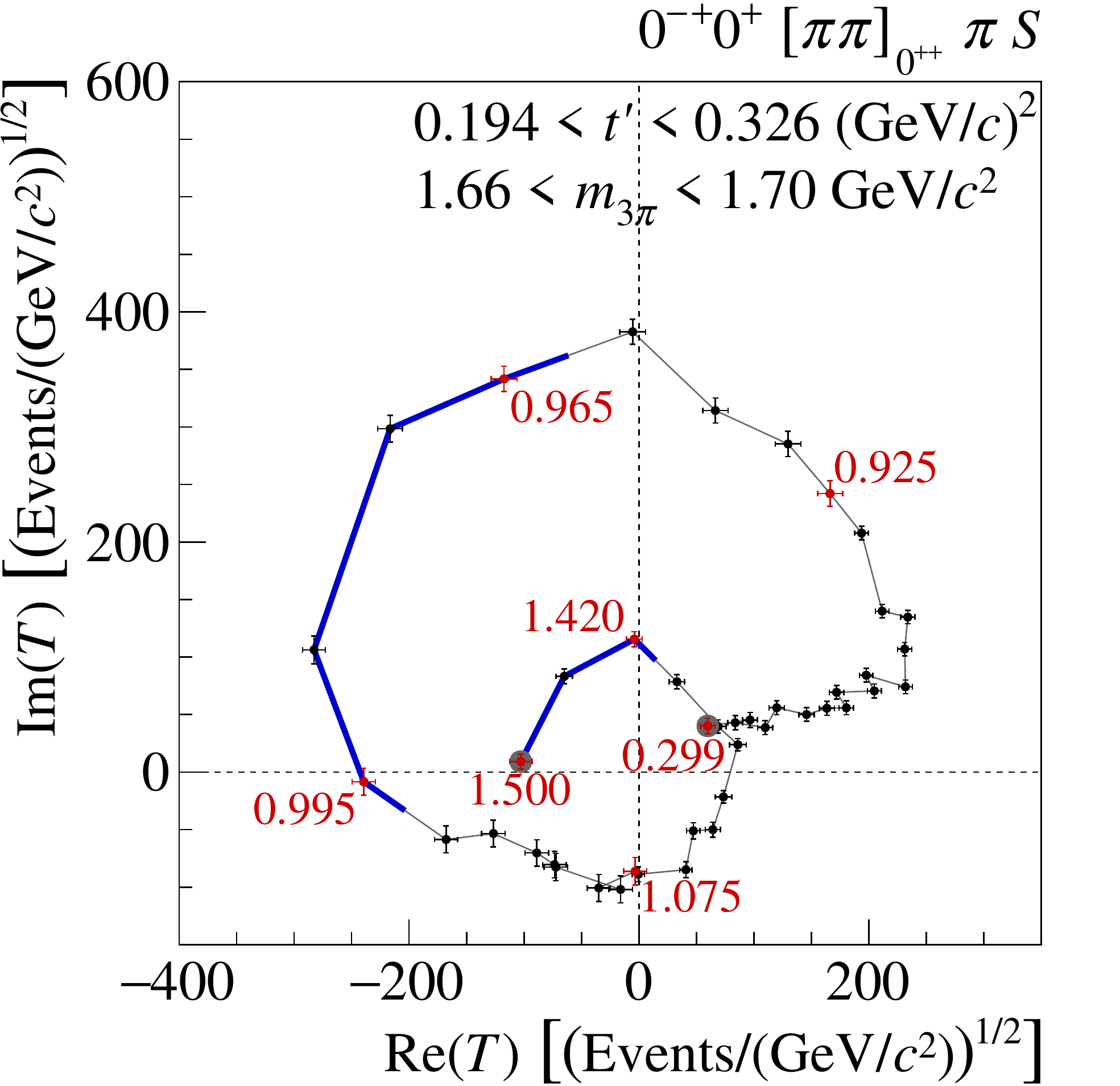}
\includegraphics[width=.33\columnwidth]{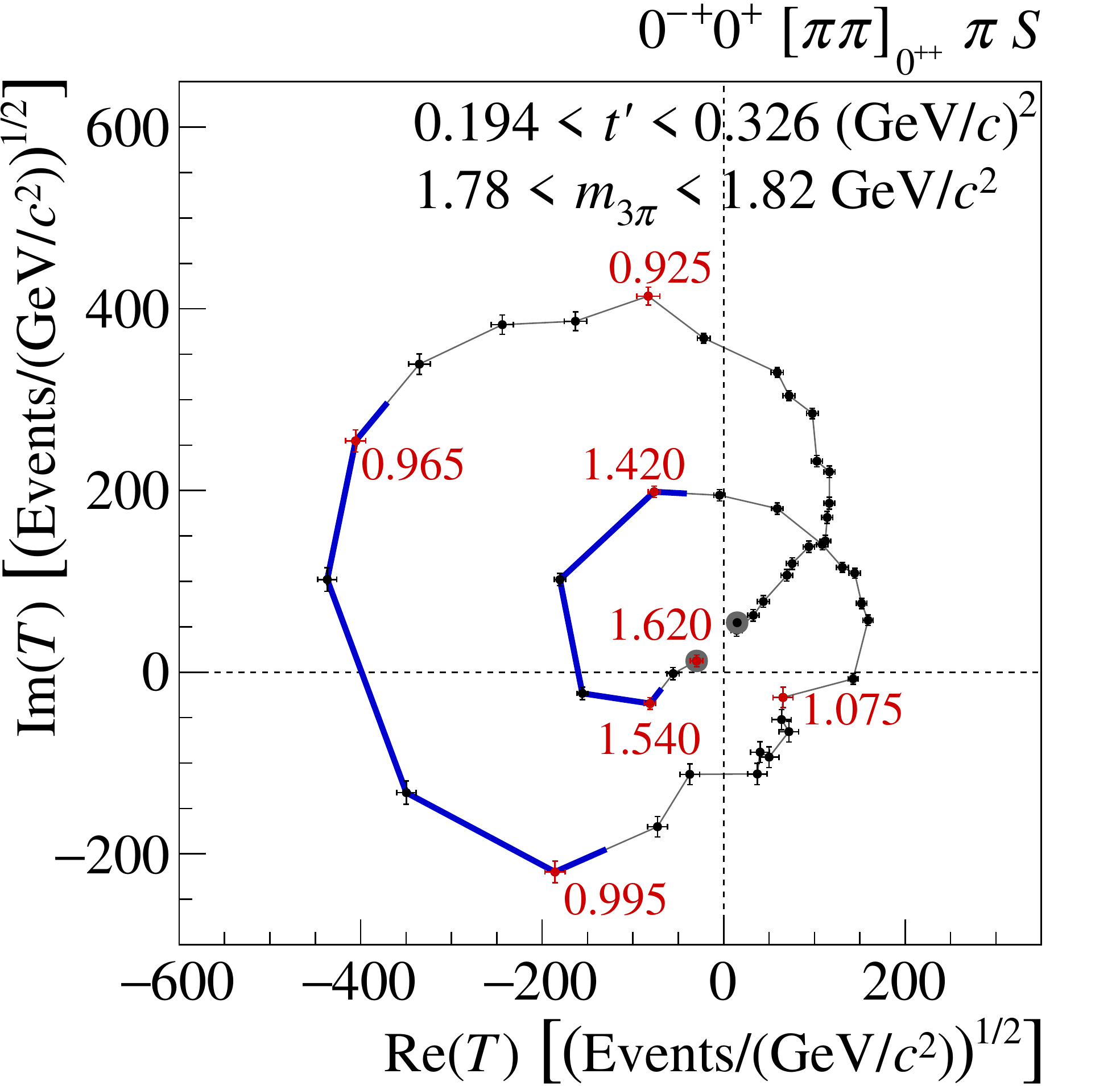}
\includegraphics[width=.33\columnwidth]{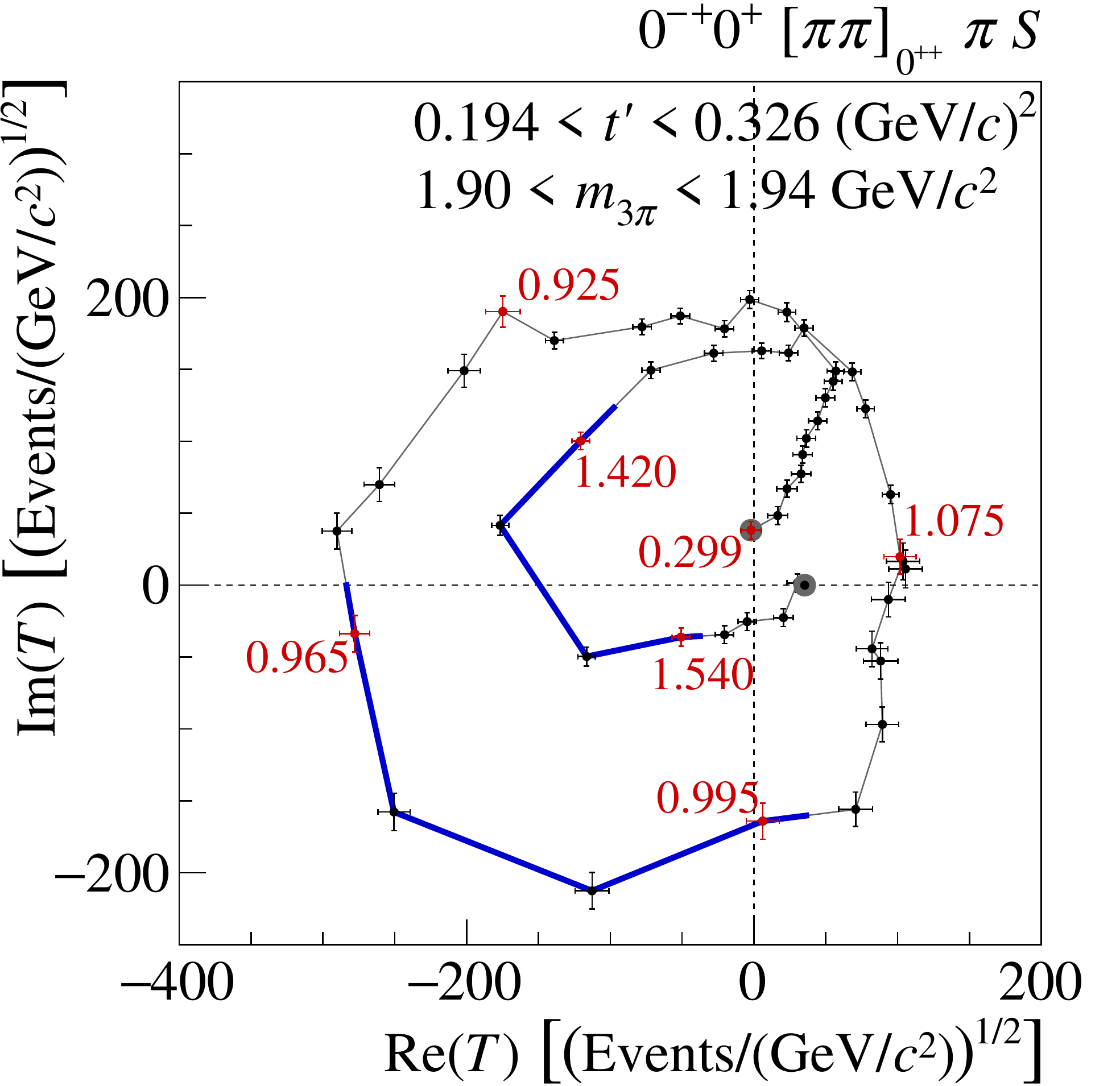}\\
\caption{Argand diagrams for the extracted $\pipi$ amplitude, below (left), on (middle) and above (right) the $\pi(1800)$ resonance. The phases are measured with respect to the $1^{++}0^+\rho\pi\,S$ wave. The $m_{\pi^+\pi^-}$ ranges
$0.96$--$1.00\GeV/c^2$ and $1.40$--$1.56\GeV/c^2$, corresponding to the $f_0(980)$ and the $f_0(1500)$ are highlighted in blue \cite{Adolph:2015pws}.}
\end{figure}

The first wave with freed isobars considered here is the $0^{-+}0^+[\pi\pi]_{0^{++}}\pi\,S$ wave. In this wave, three fixed isobars of the conventional PWA were absorbed: the $f_0(500)$, the $f_0(980)$, and the $f_0(1500)$. Since the result is now binned in $m_{3\pi}$ 
and $m_{\pi\pi}$, a two-dimensional picture is obtained (see Fig. \ref{fig::0mp}).\\
The most striking feature is a clear peak corresponding to the $\pi(1800)$ decaying to $f_0(980)\,\pi^-$. A smaller peak corresponding to $\pi(1800)\to f_0(1500)\,\pi^-$ is also visible. At low $2\pi$ and $3\pi$ masses, 
some structures are visible which are believed to be of non-resonant origin since they move with $t^\prime$.\\
In Fig. \ref{fig::0mpInt} the intensity distributions are shown for three values of $m_{3\pi}$ below, on, and above the $\pi(1800)$ resonance. A clear $f_0(980)$-peak shows in every $m_{3\pi}$-bin. Its magnitude
follows the intensity peak of the decaying $\pi(1800)$. A broad structure can also be observed as well as a second small peak corresponding to the $f_0(1500)$.\\
The two $f_0$ resonances can also be seen in the corresponding Argand diagrams in Fig. \ref{fig::0mpArg} which show two semi-circular structures corresponding to the $f_0(980)$ and the $f_0(1500)$ resonances. 
An overall rotation of the diagrams with increasing $m_{3\pi}$ can be seen, coming from the phase motion of the decaying $\pi(1800)$.

\subsection{\boldmath$1^{++}0^+[\pi\pi]_{0^{++}}\pi\,P$}
\begin{figure}[h]
\label{fig::1pp}
\includegraphics[width=\widthTwo]{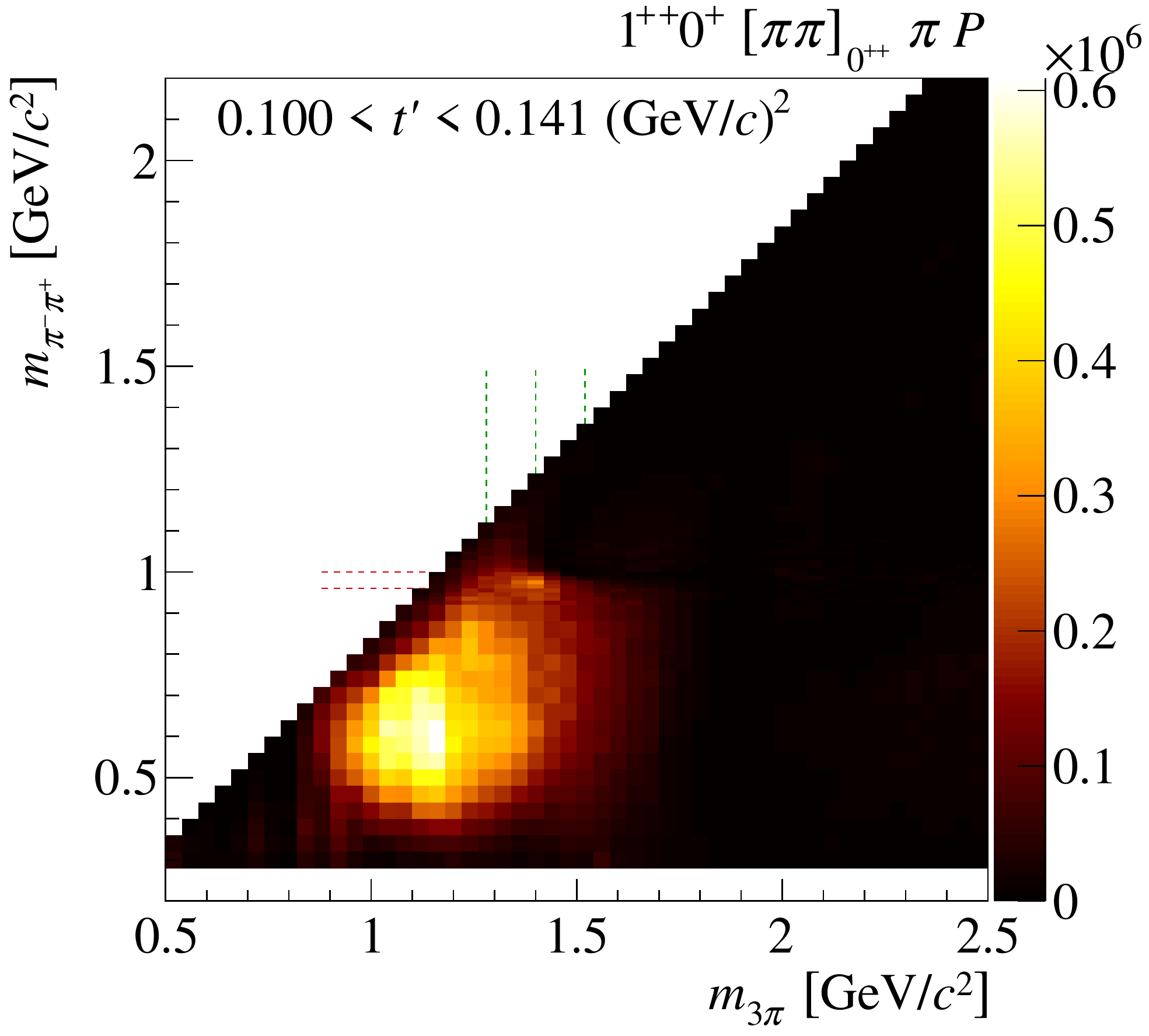}
\includegraphics[width=\widthTwo]{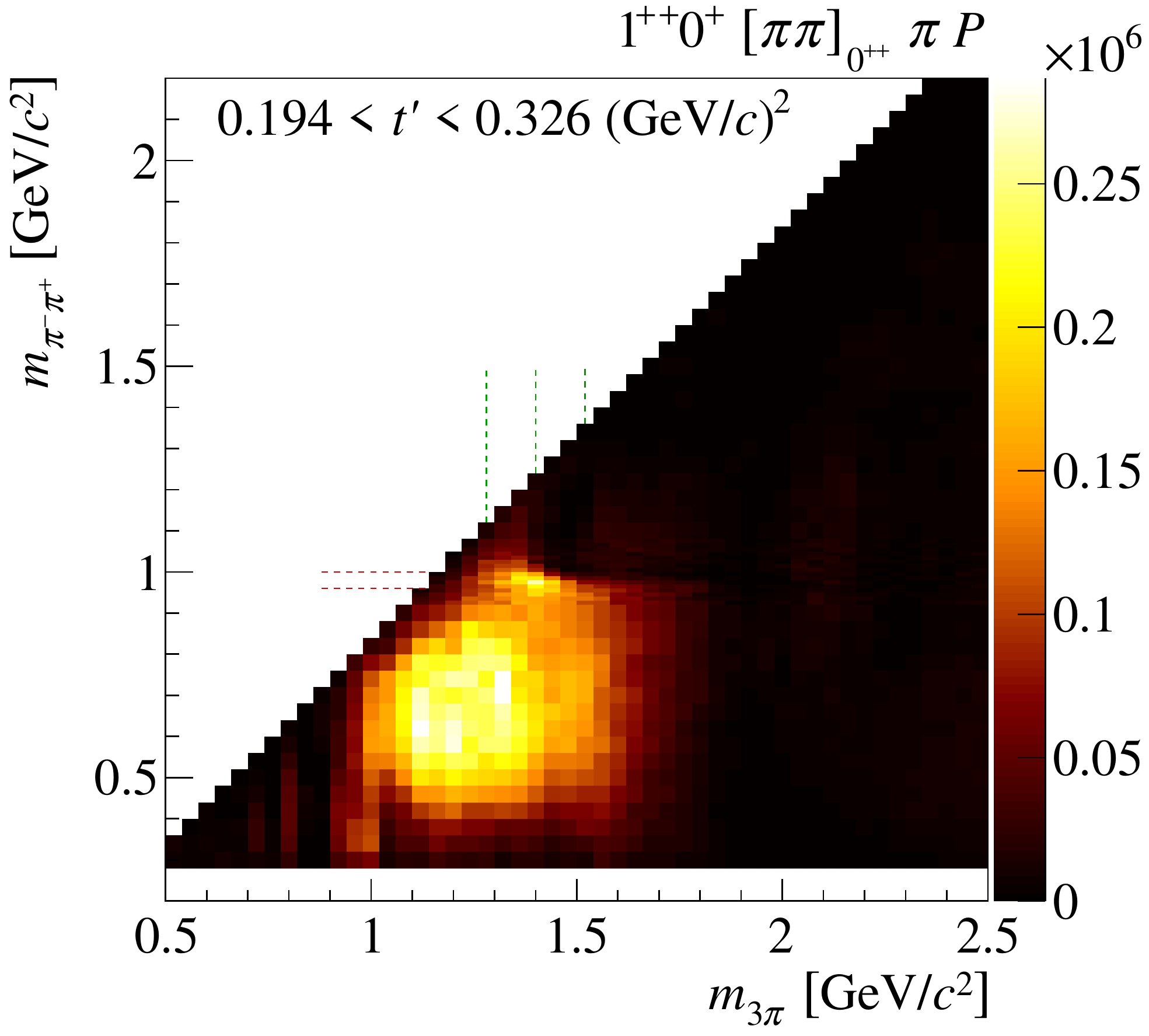}
\caption{Two-dimensional intensity distribution of the $1^{++}0^+[\pi\pi]_{0^{++}}\pi\,P$ wave depending on $m_{3\pi}$ and $m_{\pipi}$ for two different bins of the four-momentum transfer $t^\prime$ \cite{Adolph:2015pws}.}
\end{figure}

\begin{figure}[h]
\label{fig::1ppInt}
\includegraphics[width=.33\columnwidth]{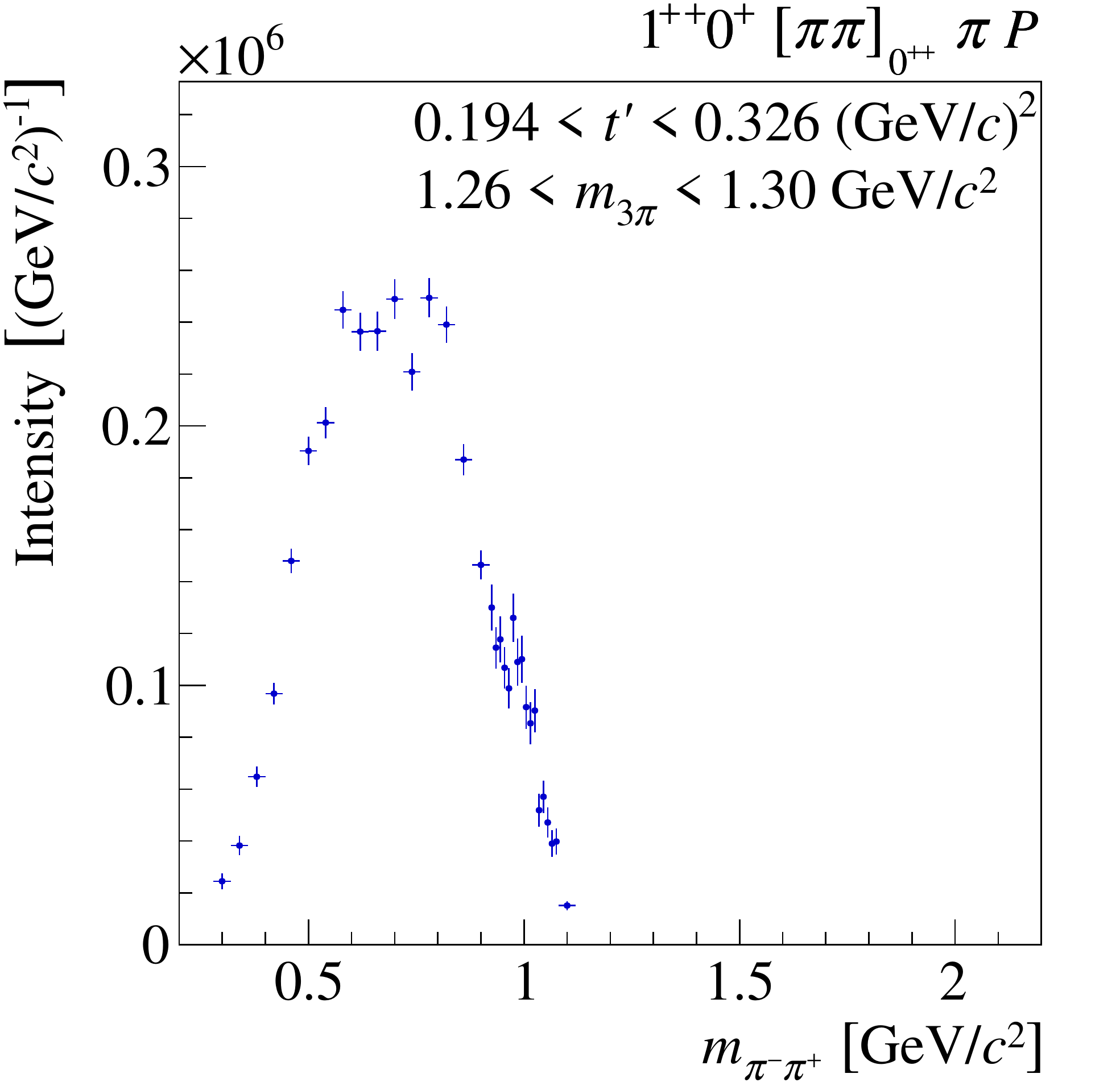}
\includegraphics[width=.33\columnwidth]{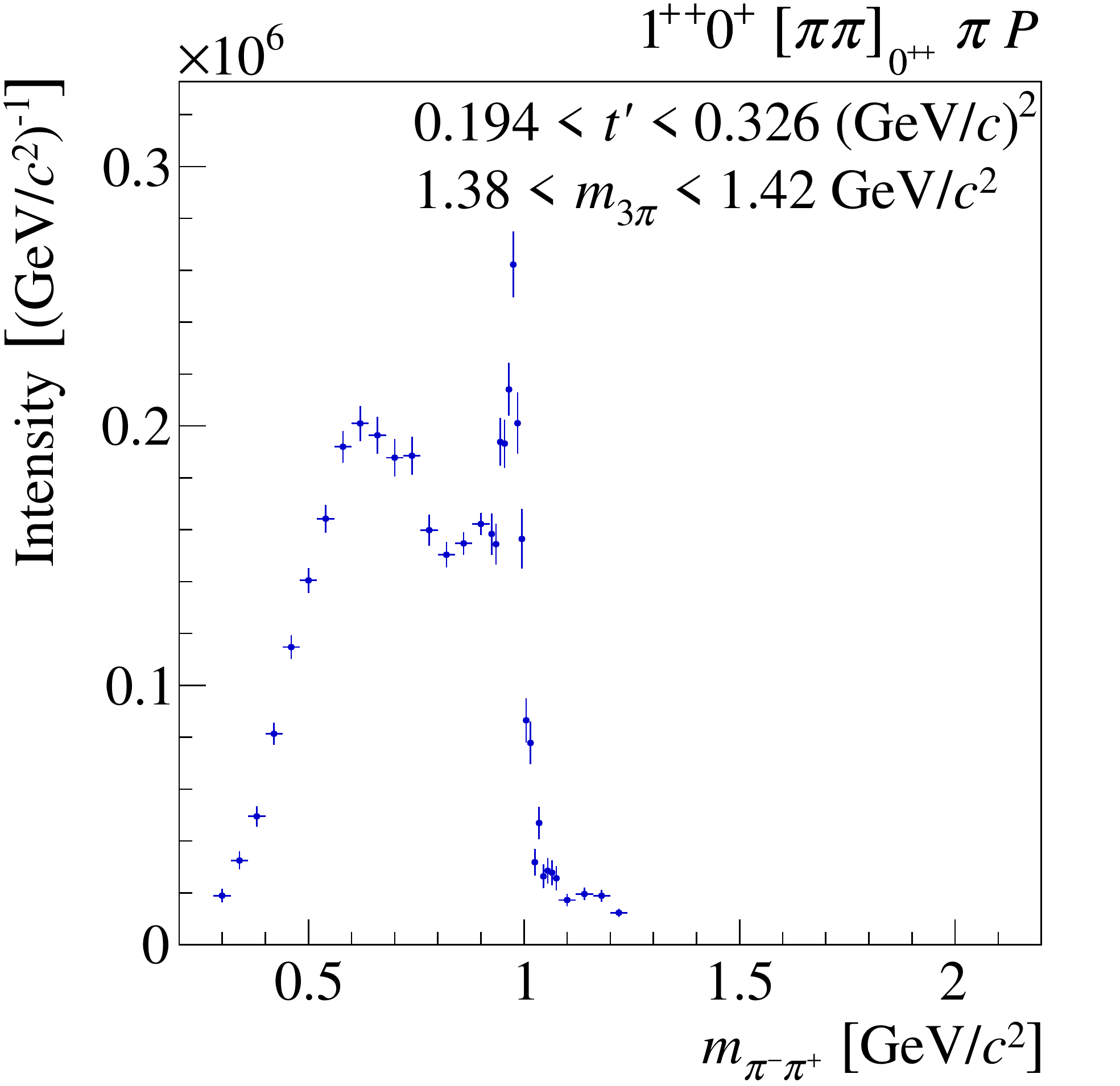}
\includegraphics[width=.33\columnwidth]{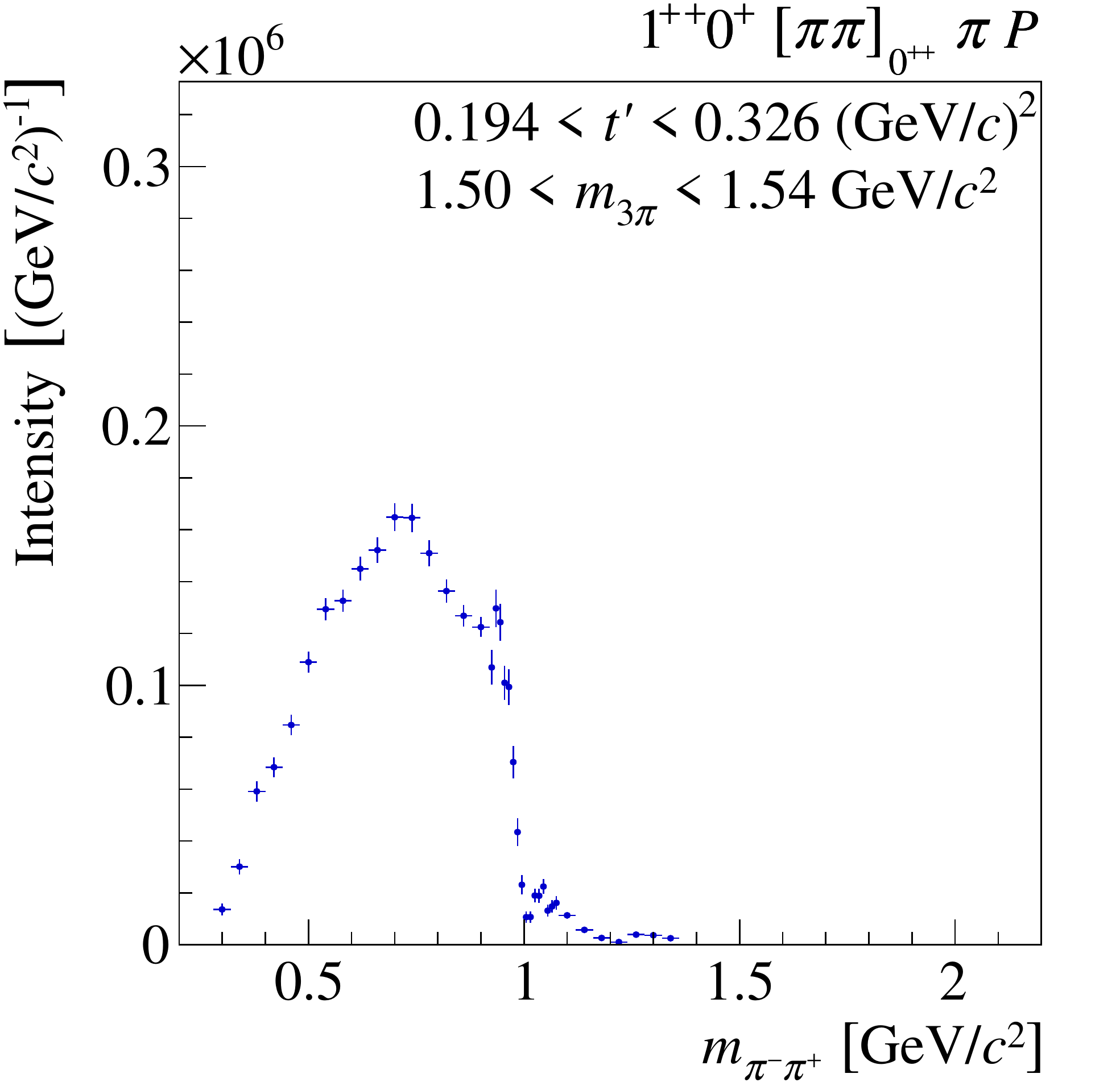}\\
\caption{Intensity distribution of the $1^{++}0^+[\pipi]_{0^{++}}\pi\,P$ amplitude at $m_{3\pi}$ bins below, on and above the $a_1(1420)$ resonance \cite{Adolph:2015pws}.}
\end{figure}

\noindent The second freed wave is the $1^{++}0^+[\pi\pi]_{0^{++}}\pi\,P$ wave, which absorbs only two isobaric waves, since no $f_0(1500)$ wave was included in the conventional analysis.\\
The corresponding two-dimensional intensity distribution is shown in Fig. \ref{fig::1pp}. The dominant feature is a very broad structure at low $2\pi$ and $3\pi$ masses. Looking at the same distribution in different regions of
 $t^\prime$ suggests a non-resonant origin of this structure, since it moves with $t^\prime$. A second feature is a narrow peak at $m_{3\pi} \approx 1.4\,\GeV/c^2$ and a $m_{\pipi} \approx 0.98\,\GeV/c^2$. 
This confirms that the new structure, named $a_1(1420)$, is not an artifact of the used $f_0(980)$ parametrization.\\
Fig. \ref{fig::1ppInt} depicts the $\pi^+\pi^-$ intensity distributions around this new resonance. It can be seen that the $f_0(980)$ peak is only visible directly at the $a_1(1420)$ peak position, 
while below and above only the broad structure remains.\\
\subsection{\boldmath$2^{-+}0^+[\pi\pi]_{0^{++}}\pi\,D$}
\begin{figure}[h]
\label{fig::2mp}
\includegraphics[width=\widthTwo]{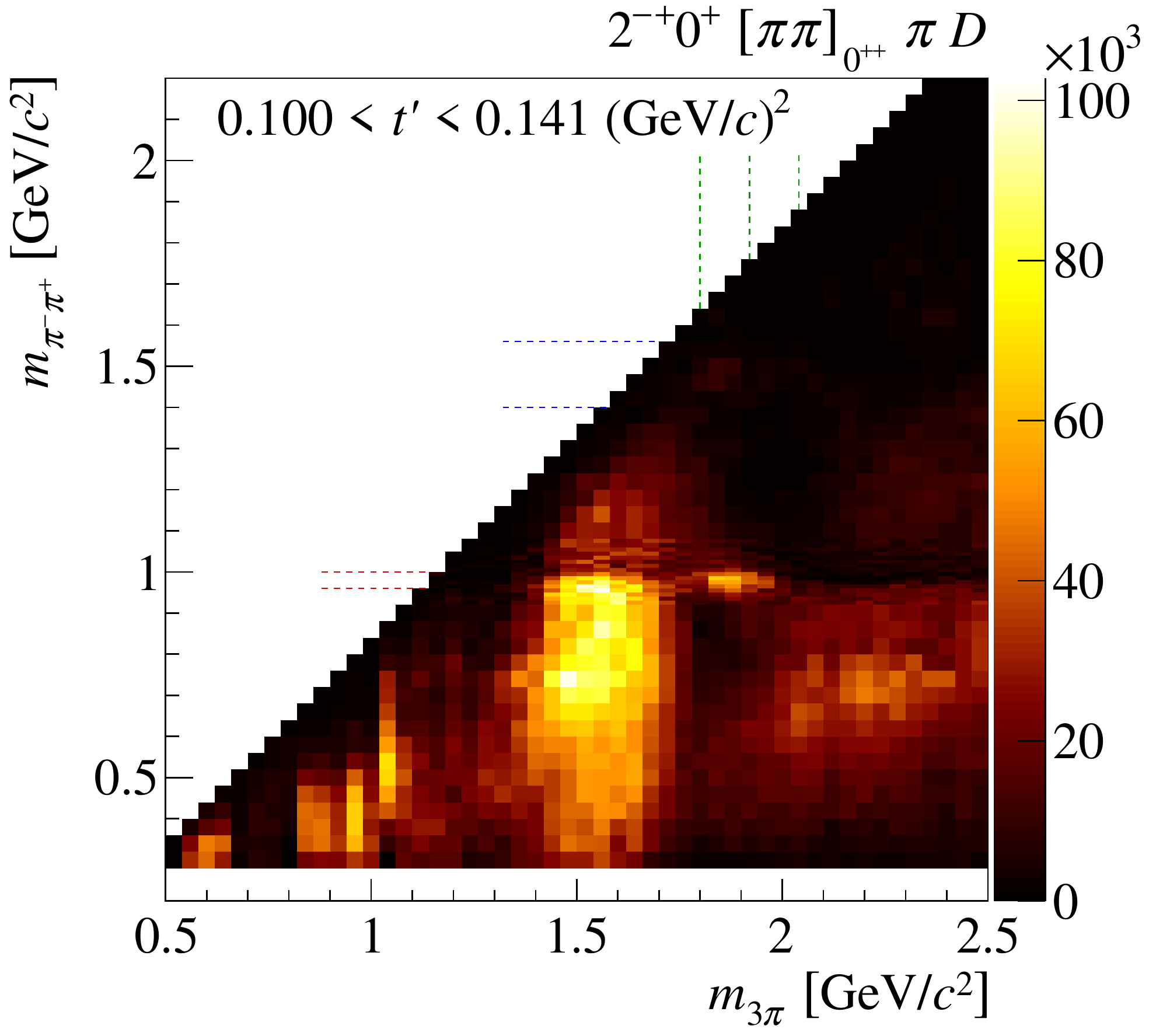}
\includegraphics[width=\widthTwo]{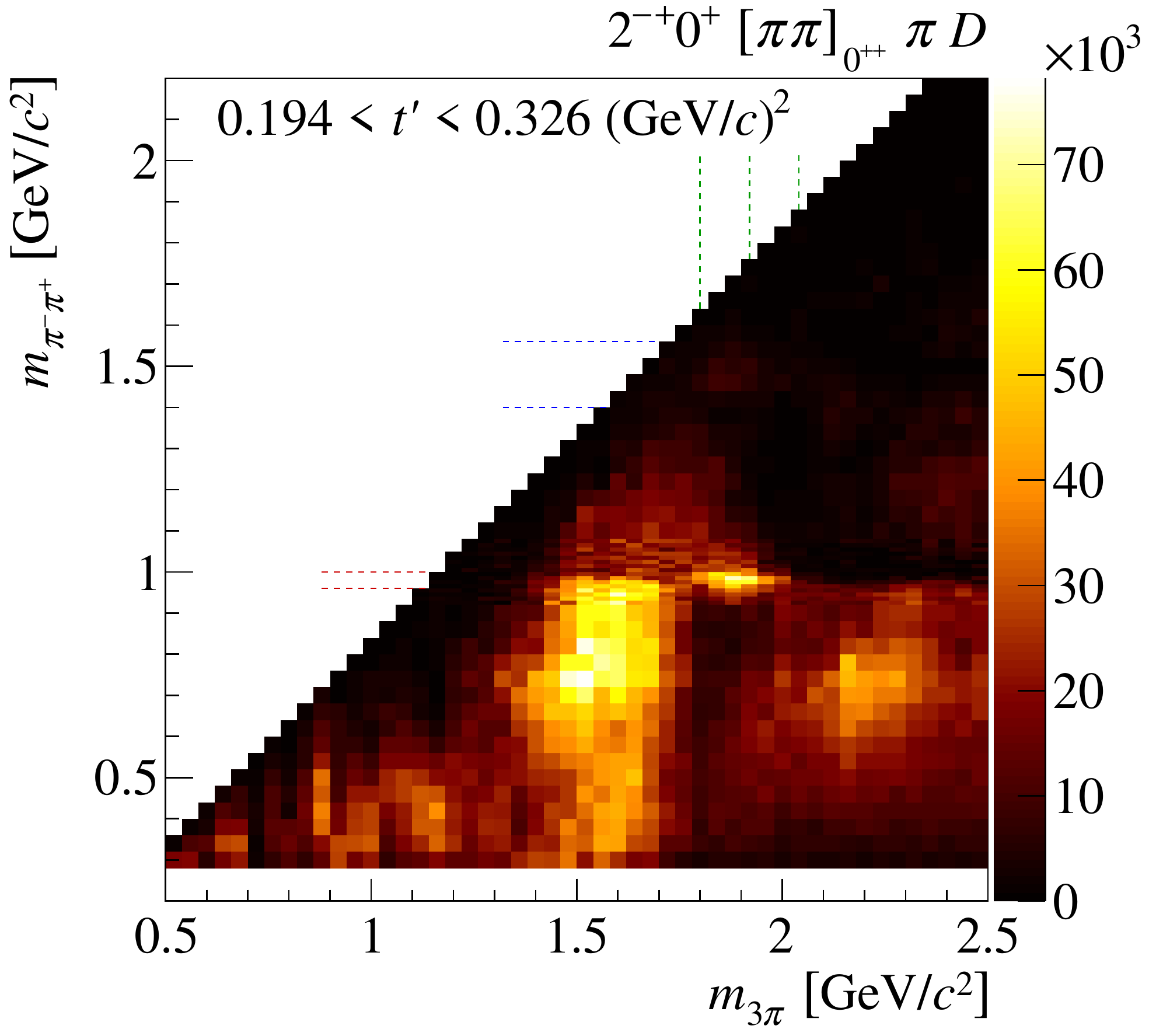}
\caption{Left: Two-dimensional intensity distribution for the freed $2^{-+}0^+[\pi\pi]_{0^{++}}\pi\,D$ wave for two different bins of the four-momentum transfer $t^\prime$ \cite{Adolph:2015pws}.}
\end{figure}

\noindent The third freed wave is the $2^{-+}0^+[\pi\pi]_{0^{++}}\pi\,D$ wave. Here again a two dimensional picture was obtained, which is shown in Fig. \ref{fig::2mp}. A peak corresponding to the $\pi_2(1880)\to f_0(980)\,\pi^-$ is visible. 
Many broad structures of unknown origin are also observed. Possible sources are non-resonant contributions as well as cross talk with fixed-isobar waves with imperfect parametrizations e.g. large waves with $J^{PC} = 2^{-+}$ 
decaying into $\rho(770)\,\pi$, where a relativistic Breit-Wigner amplitude is used to describe the $\rho(770)$.

\section{CONCLUSION}
\label{sec::conclusion}
The huge data set for the process $\pi^-p\to\pipipi p$ allows to apply a novel analysis method, which does not rely on previous knowledge on the isobar amplitudes but instead allows to extract these amplitudes from the data.\\
A first analysis of this kind was performed, freeing the isobar parametrizations of three waves with $J^{PC} = 0^{++}$ isobars. Most of the expected features are reproduced, especially the decay of the new $a_1(1420)\to f_0(980)\,\pi^-$. 
This confirms the new resonance observed in the conventional analysis. In addition to resonances, broad structures are observed, which typically change their shape with $t^\prime$ and may originate from non-resonant processes or cross talk with waves that still employ 
fixed isobar amplitudes.\\
The latter effect is currently being studied by extending the new approach to more waves. The current goal is a total of eleven freed waves, which then would describe a large fraction of the observed intensity. 
Unfortunately ambiguities in the form of linear dependent modes appear when freeing more than one wave with the same $3\pi$ quantum numbers. How to remove 
these modes is subject to current studies.

\bibliographystyle{aipnum-cp}%
\bibliography{proceedings_HADRON}%

\end{document}